\documentclass[aps,prd,preprint,groupedaddress]{revtex4}
\usepackage{graphicx}

\newcommand{\be}{\begin{equation}}
\newcommand{\ee}{\end{equation}}
\newcommand{\dd}{\text{d}}

\begin{document}

\title{Doppelg\"anger defects}


\author{Melinda Andrews}
\email[]{mgildner@sas.upenn.edu}
\author{Matt Lewandowski}
\email[]{mattlew@stanford.edu}
\author{Mark Trodden}
\email[]{trodden@physics.upenn.edu}
\author{Daniel Wesley}
\email[]{dwes@sas.upenn.edu}
%\thanks{}
\affiliation{Center for Particle Cosmology, Department of Physics and Astronomy, University of Pennsylvania, 209 S 33rd Street, Philadelphia PA 19104-6396 USA}

\date{\today}

\begin{abstract}

\noindent We study $k$-defects -- topological defects in theories with more than two 
derivatives and second-order equations of motion -- and
describe some striking ways in which these defects both resemble
and differ from their  analogues in canonical scalar field theories.
We show that, for some models, the homotopy structure of the
vacuum manifold is insufficient to establish the existence of $k$-defects, 
in contrast to the canonical case.
These results also constrain certain families of DBI 
instanton solutions in the $4$-dimensional effective theory.
We then describe a class of $k$-defect solutions, which
we dub ``doppelg\"angers", that
precisely match the field profile and energy density of their canonical
scalar field theory counterparts.
We give a complete characterization of Lagrangians 
which admit doppelg\"anger domain walls. 
By numerically computing the fluctuation eigenmodes about domain wall
solutions, we find different spectra for 
doppelg\"{a}ngers and canonical walls, allowing us to distinguish between
$k$-defects and the canonical walls they mimic.
We search for doppelg\"angers for cosmic
strings by numerically constructing solutions of DBI and 
canonical scalar field theories.
Despite investigating several examples, we are unable to find 
doppelg\"anger cosmic strings, hence the existence of doppelg\"angers
for defects with codimension $>$ 1 remains an open question. 
\end{abstract}

%\keywords{}

\maketitle

%\newpage
%\tableofcontents
\newpage

\section{Introduction\label{s:intro}}
Topological defect solutions to classical field theories have applications in
many areas in physics, and in
particular may have important implications for the evolution of the universe~\cite{vilenkinshellard,Achucarro:1999it,Nambu:1977ag,Davis:1997bs}.
In the early universe,
such defects may have formed as the universe cooled and various
gauge or global symmetry groups were broken.
Some defects, such as GUT monopoles, can lead to potential cosmological problems which historically inspired the development of the theory of cosmic inflation.
Other defects, such as cosmic strings, are potentially observable in the present day;
for example by affecting the spectrum of perturbations observed in the microwave background and matter distributions (although defects cannot play
the dominant role in structure formation). Further, the microphysics of such objects may be important in some circumstances, such as weak scale baryogenesis~\cite{Brandenberger:1988as,Brandenberger:1991dr,Brandenberger:ys,Brandenberger:1994bx}. Another interesting possibility arises if the strings
are superconducting, as originally pointed out by Witten~\cite{Witten:eb}, since the evolution of a network of such superconducting cosmic strings can differ from a nonsuperconducting one. In particular, the supercurrent along loops of string can build up as the loop radiates away its energy, affecting the endpoint of loop evolution. This supercurrent can become large
enough to destabilize the loop or may compete with the tension of the string loop
and result in stable remnants, known as {\it vortons}~\cite{Davis:ij}, with potentially important consequences for cosmology~\cite{Brandenberger:1996zp,Carter:1999an}.

In this article, we investigate new features of topological defects in scalar field theories with
non-canonical kinetic terms.  In particular, we study kinetic terms with more
than two derivatives, but which lead to second-order equations of motion.
These scalar field theories are similar to those employed in $k$-essence models
which have been studied in the context of cosmic acceleration and were
introduced in~\cite{ArmendarizPicon:2000ah,ArmendarizPicon:2000dh,ArmendarizPicon:1999rj}.
Kinetic terms of this general type also play an important role in other interesting models, such as those of
ghost condensation~\cite{ArkaniHamed:2003uy} or Galileon~\cite{Nicolis:2008in} fields.
The topological defects present in this general class of
theories are often termed ``$k$-defects," and some aspects of these objects
have been studied in earlier works~\cite{Babichev:2006cy,Babichev:2007tn,Sarangi:2007mj,Adam:2007ij,Jin:2007fz,Bazeia:2007df,Babichev:2008qv,Bazeia:2008zx,Adam:2008rf,Endlich:2010zj}.

In this work we report on some surprising aspects of $k$--defects, especially
$k$-domain walls and their associated instantons.  We find that there are very
reasonable choices for the $k$-defect kinetic term - such as the Dirac-Born-Infeld (DBI) form - for which there are no static defect solutions in a range of parameters, despite the
fact that the potential may have  multiple minima.  Thus, unlike canonical
scalar field theories, knowledge of the homotopy groups of the
vacuum manifold is sometimes
insufficient to classify the spectrum of topological defects.  Due to the
close connection between domain walls and instantons, this result also constrains certain instanton solutions
to non-canonical $4$-dimensional effective theories.  

Perhaps more surprisingly, it is also possible for $k$-defects to masquerade as canonical scalar field domain walls.  By this, we mean the following: given
a scalar field $\phi$ with canonical kinetic term and potential $V(\phi)$, then, up to rigid translations $x \to x + c$,
the field profile $\phi(x)$ and energy density $E(x)$ are uniquely
determined for a solution containing a single wall.  We show that there always exists a class of $k$-defect
Lagrangians which generate precisely the same field profile and energy density profile as the unique canonical defect.
To an observer who  measures the
field profile and energy density of the configuration, any $k$-defect in this class
precisely mimics the canonical domain wall.
Nevertheless, despite having identical defect solutions, we show that these two theories
are not reparameterizations of each other, since the fluctuation spectra
about the walls are different.

Most of our analytical work is carried out for scalar field theories with
domain wall solutions.  In order to study the generalization to other
topological defects, we carry out a numerical investigation of global
cosmic string $k$-defect solutions.
For the natural generalization of the DBI kinetic term, we show it
is possible to match either the field profile or energy density of the
canonical global string, but not both simultaneously.
Thus while we are unable to find an analogue of the
doppelg\"anger domain walls in this case,  we
cannot
conclusively show they do not exist.

This paper is organized as follows. In section \ref{s:kdefects}
we describe the general theory of $k$-defects and use the
specific example of the Dirac-Born-Infeld (DBI) action to illustrate how the question of existence of defects is more
complicated than the canonical case. We also discuss instanton solutions to the DBI action and compare our conclusions to existing discussions in the literature.
Section \ref{s:mimic} introduces the idea of doppelg\"anger domain walls, which can precisely mimic the field profile and energy density of a canonical domain wall.  We establish conditions for the existence of doppelg\"angers, and discuss the
fluctuation spectra about doppelg\"anger and canonical walls.  In Section
\ref{s:kstrings} we employ numerical methods to search for doppelg\"anger
cosmic strings, but are unsuccessful.  We conclude in Section
\ref{s:discussion}.

\section{Existence and Properties of $k$-Defects and Instantons}
\label{s:kdefects}

Our discussion focuses on two families of models involving a scalar field.
The first family consists of 
canonical scalar field theories is of the form
\be
S = \int \left[-\frac{1}{2} (\partial \phi)^2 - V(\phi)\right] \;\dd^4 x \ ,
\ee
where we use the $(-+++)$ metric signature, set
$\hbar = c = 1$, and  
define 
$(\partial \phi)^2\equiv \eta^{\mu\nu}(\partial_{\mu}\phi)\partial_{\nu}\phi$.  
Although we focus our discussion on four spacetime dimensions, 
essentially all of our conclusions regarding domain walls 
apply in any spacetime dimension $> 2$, since all but one of the
spatial dimensions are spectators.

The second family of models   
generalizes  the canonical scalar field 
theory by including additional derivatives of $\phi$.  This family is 
described by actions of the form
\be\label{e:PX}
S = \int \left[P(X) - V(\phi)\right] \;\dd^4 x \ ,
\ee
where we define
\be
X = ( \partial \phi )^2 = - \dot\phi^2 + (\nabla\phi)^2 \ .
\ee
We refer to a Lagrangian of the form (\ref{e:PX}) as a ``$P(X)$ Lagrangian."
(Note that there are multiple conventions for the definition of 
$X$ in the literature).  The 
canonical scalar field theory corresponds to $P(X) = -X/2$.  While there
are more than two derivatives of $\phi$ in the Lagrangian, by assuming that
the Lagrangian depends only on $X$ and $\phi$ as in
(\ref{e:PX}) we guarantee that the resulting equations
of motion are second order.

In this section, we show that static domain walls need not exist for all parameter ranges of a wide
variety of $P(X)$ theories,
even when the potential in (\ref{e:PX}) possesses multiple
disconnected minima.  We demonstrate this result using a specific form of $P(X)$,
corresponding to the Dirac-Born-Infeld (DBI) kinetic term.
We then adapt these results to study the properties of 
Coleman-de Luccia-type instantons in $4$-dimensional effective theories with DBI kinetic terms.

\subsection{Domain walls in naive DBI\label{ss:naiveDBI}}

A simple and well-motivated form of $P(X)$ is contained in the DBI action, given by
\be\label{e:DBI}
P(X) = M^4 - M^2 \sqrt{ M^4 + (\partial\phi)^2 } \ ,
\label{DBIkinetic}
\ee
where $M$ is a mass scale associated with the kinetic term, which we will refer
to as the ``DBI mass scale."  When
$(\partial\phi)^2 \ll M^4$, this kinetic term reduces to the canonical one.
In what follows, we set $M=1$, and hence normalize all mass scales to 
the DBI mass scale.  A kinetic term of the form (\ref{e:DBI}) can arise 
naturally
in various ways: for example, it is the four-dimensional effective theory
describing the motion of a brane with position $\phi$ in an extra dimension.
Often these kinetic terms appear along with additional functions
of $\phi$, known as ``warp factors."  These do not influence our conclusions 
in an essential way and so, for now, we will use the simple form (\ref{e:DBI}) to 
illustrate our conclusions, and return to the case with warp factors in 
Section \ref{ss:instantons}.  

We refer to the $P(X)$ Lagrangian defined
by (\ref{e:DBI}) as the ``naive" DBI theory since one is
merely adding a potential
function $V(\phi)$ to the DBI kinetic term~(\ref{DBIkinetic}).  There are other, and in some 
respects better, ways to generalize a pure DBI term and include interactions.
We will discuss one such extension extensively in Section \ref{s:mimic}.
Nonetheless, the $P(X)$ Lagrangian defined by (\ref{e:DBI}) is commonly
employed in the literature, and will provide an instructive example of $k$-defects
possessing a number of interesting properties, as we now discuss.

\subsubsection{The canonical wall}\label{s:canonicalWall}

As a warm-up, we first study the canonical domain wall profile.  We assume
that all fields depend on only one spatial coordinate $z$, and are independent
of time.   
With these assumptions, there exists 
a conserved quantity $J$ with $\dd J/\dd z = 0$, defined by
\be\label{e:canonJ}
J = \phi' \frac{\partial L}{\partial \phi'} - L = -\frac{1}{2}\phi'^2 + V(\phi) \ ,
\ee
where $L$ is the Lagrangian density.  We assume that the potential is 
positive semidefinite and  has discrete
zero-energy minima at $\phi = \phi_{\pm}$, 
such that $V(\phi_\pm) = 0$, with $\phi_- < \phi_+$.
Assuming boundary conditions where
$\phi = \phi_\pm$ at $z=\pm\infty$, we have that $V = \phi' = 0$ at $z=\pm\infty$.
Therefore $J = 0 $ at infinity, and since it is conserved, it vanishes 
everywhere.  This implies that (\ref{e:canonJ}) can be rewritten as
\be\label{e:canon1I}
\phi'^2 = 2 V(\phi) \ ,
\ee
which can be straightforwardly integrated to yield the usual domain wall
solution.  

To compute the energy density of the solution, we use the fact that
\be
H = \dot\phi  \frac{\partial L}{\partial \dot\phi} - L = -L \ ,
\ee
where $H$ is the Hamiltonian density and the second equality follows from our
assumption that the configuration is static.  Using (\ref{e:canon1I}) we
have that the energy density $E(\phi)$ is given by
\be\label{e:canonE}
H = E(\phi) = 2 V(\phi) \ .
\ee
In general, the energy density cannot be expressed
as a function of the field only, but must include the gradient.
A relation like (\ref{e:canonE}) is only true because we have a
conserved quantity for static configurations, which relates the field value
and its gradient. 
Thus, all of the physics of the static canonical domain wall is encoded in
the conserved quantity $J$.

\subsubsection{The naive DBI wall\label{s:nogo}}

We can carry out a similar derivation for the DBI wall in a $P(X)$ theory
defined by (\ref{e:PX}) and (\ref{e:DBI}).  Recalling we have set $M = 1$, the conserved quantity $J$ is
given by 
\be\label{e:DBIJ}
J = \frac{1}{\sqrt{1+\phi'^2}} - 1 + V(\phi) \ .
\ee
As in the canonical case described in Section \ref{s:canonicalWall}, 
we assume that the potential is 
positive semidefinite and  has discrete
zero-energy minima at $\phi = \phi_{\pm}$, 
such that $V(\phi_\pm) = 0$,
with
$\phi_- < \phi_+$.
We also assume the same boundary conditions, so that 
$\phi = \phi_\pm$ at $z=\pm\infty$.  Since $V = \phi' = 0$ at $z=\pm \infty$,
$J$ must vanish everywhere.
Hence, inverting (\ref{e:DBIJ}) yields
\be\label{e:DBI1I}
\phi'^2 = \frac{1}{\left[ 1 - V(\phi) \right]^2} - 1 \ .
\ee
This expression is the analogue of (\ref{e:canon1I}), and can be integrated
to give the field profile once $V(\phi)$ is specified.  Given a static
configuration, the energy density is then given by
\be\label{e:DBIE}
E(\phi)  = \frac{V(\phi) \left[ 2 - V(\phi) \right]}{1-V(\phi)} \ ,
\ee 
where we have used (\ref{e:DBIJ}) and the fact that $J=0$ everywhere.

Unlike the canonical case, it is apparent that
problems may arise when integrating (\ref{e:DBI1I}).  In the canonical case, so long
as $V(\phi)$ is bounded for $\phi \in [\phi_-,\phi_+]$, we  have
$\phi'$ finite everywhere.  This is no longer the case with (\ref{e:DBI1I}).
If there is any $\phi_1 \in [\phi_-,\phi_+]$ such that
$V(\phi_1) > 1$, then (\ref{e:DBI1I}) implies that $\phi'$ is undefined.
The problem can be traced back to (\ref{e:DBIJ}), in which the first two terms on 
the right-hand side can sum to any number between zero (when $\phi'$ vanishes) and $-1$ (when $|\phi'|$ is infinite).  Thus, at any point
where $V(\phi) > 1$, there is simply no value of $\phi'$ which will allow
the requirement that $J=0$ everywhere to be satisfied.  We conclude that
there are no nontrivial static solutions to the theory defined by
(\ref{e:DBI}) if $V(\phi) > 1$ at any $\phi \in [\phi_-,\phi_+]$.

To study the nature of the singularity, suppose we have integrated
(\ref{e:DBI1I}) from $\phi = \phi_-$ at $z = -\infty$ and have encountered
a value  $\phi = \phi_1$ at which $V(\phi_1) = 1$.  Assume that this
value is reached at $z = z_1$.
For a generic function $V(\phi)$ we have
\be
V(\phi_1 + \Delta\phi ) = 1 + v' \Delta\phi + \mathcal{O}(\Delta\phi^2) \ ,
\ee
where $v' = V'(\phi)|_{\phi=\phi_1}$.  Retaining only terms up to first
order in $\Delta\phi$ and using  (\ref{e:DBI1I}) leads to 
\be
\phi' = - \frac{1}{v' \Delta\phi} \ ,
\ee
which has the solution
\be
\phi(z) = \phi_1 + \sqrt{- \frac{2 (z-z_1)}{v'}} \ .
\ee
Hence, $\phi$ is well-defined when $z < z_1$, before the singularity is reached.
It is not defined for $z > z_1$, and at $z = z_1$ there is cusp-type singularity
in the field, at which the field value is finite but the gradient and 
all higher derivatives become infinite.

It is natural to ask whether this singularity is integrable; that is,
whether the solution can be continued past the singular point at $z=z_1$.
We now show that the solution cannot be continued, and hence
there are no global solutions to (\ref{e:DBI}) with the desired boundary
conditions.  We prove this claim for the simple case in which there
is only one connected interval of field space between the minima
for which $V(\phi) > 1$ (the generalization to the case where there are 
multiple disconnected regions where $V(\phi) > 1$ is straightforward).

The relevant region of field space 
is naturally dividied into
three intervals
\begin{eqnarray}
I_-  &\equiv& [\phi_-,\phi_1) \ , \nonumber \\
I_0 &\equiv& (\phi_1,\phi_2) \ , \nonumber \\
I_+ &\equiv& (\phi_2,\phi_+] \ . \nonumber
\end{eqnarray}
The intervals $I_\pm$ include the minima
of $V(\phi)$ and all field values for which $V(\phi) < 1$.  The interval
$I_0$ includes the field values for which $V(\phi) > 1$.  At the boundary
points $\phi_1$ and $\phi_2$ of $I_0$, $V(\phi) = 1$ and $\phi'$ reaches
$\pm \infty$.  We have shown that solutions of (\ref{e:DBI}) with the
desired boundary conditions can be constructed 
which take values in $I_\pm$, but now claim that
these solutions cannot be continued into $I_0$.

The key to proving our claim is to employ the quantity $J$, 
which must be conserved by the equations of motion, and is well-defined 
for any value of $\phi'$ (even $\phi' = \pm \infty$).  First, suppose
that we have a candidate continuation of the solution on $I_-$ 
into $I_0$. Using this 
continuation, we
 choose any point $z_\ast$ for which $\phi(z_\ast) \in I_0$, 
and use $\phi(z_\ast)$ and $\phi'(z_\ast)$ to evaluate $J$.
Since $V(\phi) > 1$ at $z_\ast$, then by inspection of
(\ref{e:DBIJ}), we conclude that $J > 0$ at $z_\ast$.
Since $J$ is 
conserved by the equations of motion, then $J$ must assume this
same positive definite value
for all points in $I_0$. Inspection of (\ref{e:DBIJ}) reveals that,
when $J > 0$, $\phi'$ is finite when $V(\phi) = 1$.  Hence, if we approach
$\phi_1$ while remaining in $I_0$, then the limiting value of $\phi'$ at
$\phi_1$ is finite.  On the other hand, we have already shown that
$J=0$ in $I_-$, and when $J=0$ we have that $\phi' = \pm\infty$ when
$V(\phi) = 1$.  Thus if we approach $\phi_1$ while remaining in $I_-$ we have
$\phi' = \pm \infty$ at $\phi = \phi_1$.

Thus, if there were  a global solution, then $\phi'$ would approach a finite
value from one side  of $\phi_1$, and an infinite value from the other
side.  This means that the purported global 
solution would not match smoothly across the 
singularity at $\phi_1$; a contradiction.  Hence we conclude 
that global solutions do not exist.

While the above statements are strictly correct within the context of the specific Lagrangian we have used, there are potential problems in treating
the DBI Lagrangian as an effective field theory near the singularity
at $\phi_1$.  Expanding the Lagrangian $L$ about 
a static background solution $\phi(z)$
 gives terms of the form 
\be
\delta_2 L \supset - \frac{\delta\phi'(z)^2}{2 \left(
1 + \phi'(z)^2 \right)^{3/2}}
\ee
at quadratic order in the fluctuation $\delta\phi(z)$.  Hence 
the kinetic term for fluctuations vanishes
as we 
approach the point $z_1$ where $\phi = \phi_1$ and $\phi'(z) \to \infty$.
Near the singularity, 
the effective theory is strongly coupled, 
quantum corrections to (\ref{e:DBI}) are large, and the precise
functional form of (\ref{e:DBI}) is not trustworthy.  Whether these
corrections invalidate our conclusions is an open question.
Nonetheless, our analysis shows that the topological structure of the vacuum is not enough to guarantee the existence of topological defects in models
with extra derivatives.

\subsection{Application to instantons\label{ss:instantons}}

Domain wall solutions are closely related to the solutions to Euclidean
field theories employed in constructing instantons.  This is because the
lowest-energy Euclidean configurations typically depend on a single 
coordinate, and thus have
 essentially the same structure as domain wall solutions.
Although there are some differences, the correspondence
becomes exact in the thin-wall limit.  For example, to study the Coleman-de Luccia instanton occurring
in a canonical field theory one considers the Euclidean action
\be\label{e:canonSE}
S_E = 2 \pi^2 \int \left[ \frac{1}{2} \phi'^2  + V(\phi) \right] \,  \rho^3 \, \dd \rho \ ,
\ee
where $\rho$ is the Euclidean radial coordinate, and in this subsection only
we take $\phi' \equiv\partial \phi / \partial \rho$.  Instantons are
solutions of the equations
of motion of this action.  The main difference between the action (\ref{e:canonSE}) and the canonical domain wall action is the
presence of the $\rho^3$ factor in the integration measure.  When the
thickness of the wall is much smaller than $\rho$ -- the ``thin wall limit" --
the measure factor can be neglected, and the instanton problem reduces to 
the domain wall problem. Thanks to this correspondence, we
can apply some of our domain wall techniques to the study of instantons
in higher derivative theories.  

The properties of instanton
solutions for  DBI actions of the form (\ref{e:DBI}) have been studied
previously.  In particular, in \cite{Brown:2007zzh} a generalization of 
(\ref{e:DBI}) was considered, of the form
\be
S = \int \left[f(\phi)^{-1} \left(  1 - 
 \sqrt{ 1 + f(\phi)(\partial\phi)^2 } \right) - V(\phi)\right] \; \dd^4 x \ ,
\ee
where the function $f(\phi)$ is the ``warp factor."  The corresponding Euclidean action  is
\be\label{e:SEwarp}
S_E = 2\pi^2 \int \left[ f(\phi)^{-1} \left( -1 + \sqrt{1 + f(\phi)\phi'^2} \right) + V(\phi) \right] \,  \rho^3 \, \dd \rho \ .
\ee
The authors of \cite{Brown:2007zzh} pointed out that
solutions for $\phi$ develop cusp-like
behavior once $V(\phi)$ became large.  It was argued that this corresponded
to instantons where the field profile is multi-valued, and the
graph of $(z,\phi(z))$ traces out an S-curve, as illustrated in Figure 2
of \cite{Brown:2007zzh,Brown:2007vha}.
  Geometrically,
if $\phi$ is interpreted as the position of a brane in an extra dimension,
this would correspond to the brane 
doubling back upon itself.
However, if we treat the action~(\ref{e:SEwarp}) as a $4$-dimensional effective theory, then,  as we shall explain below,
these solutions only exist for special choices of the functions
$f(\phi)$ and $V(\phi)$.

To apply our previous results, we must
generalize them to include
 the measure factor and the warp factor.
Since we are concerned entirely with
the Euclidean equations of motion arising from (\ref{e:SEwarp}), which are
not affected by constants multiplying the Lagrangian, it is convenient
to absorb a factor of $-2\pi^2$ into $S_E$, and thus consider the
Euclidean Lagrangian
\be
L_E = \left[ f(\phi)^{-1} \left( 1 - \sqrt{1 + f(\phi)\phi'^2} \right) - V(\phi) \right] \rho^3 \equiv \hat{L}_E \rho^3 \ .
\ee
The Lagrangian $L_E$ incorporates the effects of the warp factor and
the measure factor, while
$\hat{L}_E$ incorporates warp factor effects alone.  For
static solutions, the conserved 
quantity corresponding to $\hat{L}_E$ is  
\be\label{e:Jhat}
\hat{J} = f(\phi)^{-1} \left[
\frac{1}{\sqrt{1+ \phi'^2 f(\phi)}} - 1 + V(\phi) 
\right] \ ,
\ee
which may be compared to (\ref{e:DBIJ}).  It is important to stress that
(\ref{e:Jhat}) is not precisely conserved: $\hat{J}$ arises from 
$\hat{L}_E$, whereas the full equations of motion arise from $L_E$, which 
contains the measure factor $\rho^3$.  
The full equations of motion imply that
\be
\frac{\partial \hat{J}}{\partial z} = \left(\frac{3}{\rho}\right)
\frac{\phi'^2}{\sqrt{1 + f(\phi) \phi'^2}} \ ,
\ee
and this
non-conservation of $\hat{J}$ describes important physics. Just as in the canonical instanton, this is what enables 
tunneling between minima of $V(\phi)$ with different vacuum energies, an
essential feature of the Coleman-de Luccia instanton.
However, in the thin-wall limit, where the width of the instanton solution is
much less than $\rho$, the total change in $\hat{J}$ will be very small
across the instanton wall.  Hence, if we focus only on the instanton wall
itself, $\hat{J}$ is effectively conserved.  

The
approximate conservation of $\hat{J}$  enables us to employ
some of our domain wall techniques from Section \ref{s:nogo} to
the instanton problem, and to show that
there is no solution to the Euclidean equations of motion in which $\phi$
curls back on itself.  Suppose such a solution did, in fact, exist. Folding back upon itself would occur
when $\phi' = \infty$, and we denote the
value of $\phi$ at which this occurs as $\phi_\ast$, and the corresponding
value of $\rho$ by $\rho_\ast$.  Using (\ref{e:Jhat}) and working backwards,
we find this defines a value of $\hat{J}$ given by
\be\label{e:Jhatstar}
\hat{J}_\ast = \frac{V(\phi_\ast) - 1}{f(\phi_\ast)} \ .
\ee
Approximate conservation of $\hat{J}$ 
means that we can take $\hat{J} = \hat{J}_\ast$ when dealing with physics
in the vicinity of the wall.  Despite the fact that the point $\phi = \phi_\ast$ is in some sense singular, $\hat{J}$ must be the same on either side of
this point.  This is because, clearly, $\hat{J}$ is approximately 
conserved away from singular points (such as $\phi_\ast$).  If we denote
$\hat{J}_\pm$ as the value of $\hat{J}$ for $\phi < \phi_\ast$ and 
$\phi > \phi_\ast$, respectively, then the only way to ensure 
that Lim$_{\phi \to \phi_\ast^+} = \infty$ and Lim$_{\phi \to \phi_\ast^-} = \infty$ is to have $\hat{J}_+ = \hat{J}_- = \hat{J}_\ast$.

We now focus on a closed interval $I_\epsilon$ in $\phi$, of 
radius $\epsilon$, and
centered on $\phi=\phi_\ast$, so
$I_\epsilon = [\phi_\ast-\epsilon,\phi_\ast+\epsilon]$.  
Assuming $f(\phi)$ is smooth,
given any $\delta >0$ we can choose $\epsilon >0$ so that

\be
\frac{1}{f(\phi)\sqrt{1+ \phi'^2 f(\phi)}} \le \delta
\qquad \forall \;\; \phi \in I_\epsilon \ .
\ee 
Conservation of $\hat{J}$ then implies
\be
\Bigg{|} \frac{V(\phi) - 1}{f(\phi)} - \hat{J}_\ast \Bigg{|} \le \delta
\qquad \forall \;\; \phi \in I \ .
\ee
Using the definition (\ref{e:Jhatstar}) and taking  the $\delta \to 0$
limit, we can rewrite this condition as
\be\label{e:cont}
\frac{f'(\phi_\ast)}{f(\phi_\ast)} = \frac{V'(\phi_\ast)}{V(\phi_\ast)-1} \ .
\ee
If this condition is not satisfied, it is impossible to continue 
the solution through the singular point.  Any deviation from
(\ref{e:cont}) leads to a singular solution, and no fold is possible.
For generic functions $f$ and $V$, the condition
 (\ref{e:cont}) is not satisfied, and hence the required instanton
 solutions do not exist.

To illustrate these results, we can consider the
case $f(\phi) = 1$, corresponding to the naive DBI action studied
in Section \ref{ss:naiveDBI}.  The cusp is located at $\phi=\phi_\ast$
where $V(\phi_\ast) = 1$, and hence $\hat{J}_\ast = 0$.  
In order to fold back upon itself, $\phi$ must be greater than $\phi_\ast$
on one branch of the solution, and less than $\phi_\ast$ on the other.
Hence $V(\phi) > 1$ on one branch, and $V(\phi) < 1$ on the other, for
generic $V(\phi)$.  However, from (\ref{e:Jhat}) it is clear that there is
no solution for $\phi'$ when $V(\phi) > 1$, and hence the solution cannot
be continued through the fold.  This ultimately arises because the condition
(\ref{e:cont}) cannot  be satisfied if we take $f(\phi) = 1$.

\section{Doppelg\"anger Domain Walls\label{s:mimic}}

In Section \ref{s:nogo}, we showed that domain walls in $P(X)$ theories
can be very different from those in canonical scalar field theories.  However, in
this section, we show that in a particular class of higher-derivative
theories, the walls can actually be remarkably similar to their
canonical counterparts!
Indeed, the background solution for these walls is completely indistinguishable
from the canonical wall, with the same energy density and field profile. 
As we shall see, the two solutions differ only in their fluctuation spectra.

\subsection{An Example: Masquerading DBI}

\subsubsection{Motivating the action}

Rather than diving immediately into a general  analysis, it is instructive to begin with 
a simple and physically motivated example - the DBI action.
One way of deriving the DBI kinetic term is to consider $\phi$ to be the
coordinate of an extended object in an extra-dimensional space.  Such objects
can be described by the Nambu-Goto action, which is simply their surface
area multiplied by the tension.  If we take the higher-dimensional space to have 
coordinates $X^N$ with $N = 0,...4$ then the action is
\be\label{e:NG}
S_{NG} = - T \int \sqrt{ - \det \left[ \eta_{MN}\frac{\partial X^M}{\partial x^\mu}
\frac{\partial X^N}{\partial x^\nu} \right] } \; \dd^4 x \ ,
\ee
where $T$ is the tension, and
 $\eta_{MN}$ is the metric in the full five-dimensional space, which we
 take to be Minkowskian. Taking the embedding defined by
\be
X^N = x^N, \quad N = 0,...3, \qquad X^4 = \phi(x^\mu)
\ee
leads precisely to the $P(X)$ in (\ref{e:DBI}), modulo a constant which only
serves to set the energy of the vacuum to zero.  

This extra-dimensional setup provides a useful geometrical picture for the
origin of the DBI kinetic term. However, it is not clear how the simple 
addition of a
potential $V(\phi)$, as we have done in Section \ref{s:nogo}, can be 
interpreted in this picture.  If we hew to the extra-dimensional picture, 
it would seem that any new terms we add to the DBI action should correspond to geometrical quantities, such as the surface area of the 
membrane in the higher dimensional space.  Such an approach also ensures that these
additional terms will be compatible with the coordinate reparameterization
symmetry of the action (\ref{e:NG}).

Guided by these considerations, we  study actions in which the tension $T$ is promoted to a function
of the spacetime coordinates $X^M$, so that (\ref{e:NG}) becomes 
\be\label{e:NG1}
S_{NG} = -  \int T(X) \sqrt{ - \det \left[ \eta_{MN}\frac{\partial X^M}{\partial x^\mu}
\frac{\partial X^N}{\partial x^\nu} \right] } \; \dd^4 x \ .
\ee
Descending to the four-dimensional theory, we find that such a system 
cannot be described by a $P(X)$-type Lagrangian (\ref{e:PX})
because of the 
way in which $X$ and $\phi$ are coupled.   The 
resulting action is
\be\label{e:DBI2}
S =  \int \left[1 - \left( 1+ U(\phi) \right) \sqrt{1 + (\partial \phi)^2 }\right] \; \dd^4 x \ ,
\ee
where, as in Section \ref{s:nogo}, we have set $M = 1$, where $M$ is
the mass scale associated with the DBI kinetic term.  We have also added a constant
to the Lagrangian density in order to ensure that the energy density 
vanishes when $\phi' = 0$ and $U(\phi) = 0$.  When gradients are small and
$(\partial\phi)^2 \ll M^4$,
the Lagrangian is approximately
\be
L = 1 - \left( 1+ U(\phi) \right) \sqrt{1 + (\partial \phi)^2 }
\sim \frac{1}{2} \dot\phi^2 - \frac{1}{2} (\nabla\phi)^2 - U(\phi) \ ,
\ee
and hence $U(\phi)$ is analogous to the potential in the canonical theory.
However, as we shall see below, it plays a somewhat different role in the full theory.

\subsubsection{Dirac-Born-Infeld Doppelg\"angers}

We now ready to study defect solutions corresponding to the action (\ref{e:DBI2}).  For this action, the
conserved quantity $J$ is 
\be\label{e:DBI2J}
J = \frac{1+U(\phi)}{\sqrt{1+\phi'^2}} - 1 \ .
\ee
As before, we assume that $U(\phi)$ has two discrete minima $\phi_\pm$ where
$U(\phi_\pm) = 0$ and take boundary conditions where $\phi = \phi_\pm$ at
$z=\pm \infty$.  Thanks to the boundary conditions,  $J = 0$ at infinity,
 and therefore $J$  vanishes
everywhere because it is conserved.  Inverting (\ref{e:DBI2J}) gives
\be\label{e:DBI21I}
\phi'^2 = U(\phi) \left[ U(\phi) + 2\right] \ ,
\ee
which can be integrated to find the field profile for the defect.  The
Hamiltonian
energy density of the defect is given by
\be\label{e:DBI2E}
E(\phi) = -1 + \left[ 1+ U(\phi)\right] \sqrt{1 +\phi'^2}
= U(\phi) \left[ U(\phi) + 2\right] \ ,
\ee
where in the second equality we have used the expression (\ref{e:DBI2J}) and
the fact that $J$ vanishes.

The curious properties of the doppelg\"anger walls arise from the
fact that the right-hand
sides of (\ref{e:DBI21I}) and (\ref{e:DBI2E}) are identical: the
energy density is equal to $\phi'^2$.  The only 
other case we have seen thus far with this
property was the canonical domain wall, as seen in
(\ref{e:canon1I}) and (\ref{e:canonE}).  This property was not shared 
by the naive DBI domain wall, as can be seen from (\ref{e:DBI1I}) and
(\ref{e:DBIE}).  This means that, for static solutions arising from the
action (\ref{e:DBI2}), we can define an effective potential function
$\hat{V}(\phi)$ for the DBI wall by
\be\label{e:mimicV}
\hat{V}(\phi) \equiv \frac{1}{2} U(\phi) \left[ U(\phi) + 2\right] \ .
\ee
Note that minima of $U(\phi)$ where $U(\phi) = 0$ are also minima of 
$\hat{V}(\phi)$ where $\hat{V}(\phi)=0$.
With the identification
(\ref{e:mimicV}) the equations (\ref{e:DBI21I}) and (\ref{e:DBI2E}) are precisely
the same as the analogous equations for the canonical domain wall
(\ref{e:canon1I}) and (\ref{e:canonE}), but with the substitution
$V \to \hat{V}$.  By inverting (\ref{e:mimicV}), we find
\be\label{e:mimicU}
U(\phi) = -1 + \sqrt{1 + 2 \hat{V}(\phi)} \ .
\ee
So, we  conclude with the somewhat surprising result that:

\begin{quotation}
\noindent\emph{Given a canonical scalar field theory with a
positive semi-definite
potential $V(\phi) \ge 0$ which supports
domain wall solutions, there exists a choice for
$U(\phi)$ in the DBI theory (\ref{e:DBI2}), given by setting
$\hat{V} = V$ in (\ref{e:mimicU}), which guarantees domain walls with
precisely the same field profile and energy density.
}
\end{quotation}

\noindent In the next two sub-sections, we present two pieces of evidence
which support the idea that our claim is somewhat surprising. 
First, we show that a claim of this
nature cannot be made for arbitrary theories with extra derivatives: 
generically, there is no way to choose a potential function so that 
the higher-derivative wall mimics the canonical one.  We reinforce
this argument by deriving an explicit
condition for the existence of doppelg\"anger defects.
Second, 
we numerically compute the fluctuation
spectra about the background domain wall solution, and find they are
very different for the canonical wall and the DBI one.
This shows that
the DBI theory (\ref{e:DBI2}) is not a rewriting of the canonical 
scalar field theory, despite having solutions with identical field profiles
and energy density.

\subsection{When do Doppelg\"anger Defects Exist?\label{ss:unique}}

\subsubsection{A Counter-Example - Other $P(X)$ Theories}

While we have shown that the action (\ref{e:DBI2}) possesses doppelg\"anger
solutions, this is not a generic property of theories with higher derivatives.
The $P(X)$ theory with a DBI kinetic
term studied in Section \ref{ss:naiveDBI} already provides 
one example where a $P(X)$-type theory always leads to
domain wall solutions which differ from those of a canonical field theory, with either a different field profile or a different
energy density (or both).    The DBI wall with a $P(X)$ action of the 
type (\ref{e:DBI}) can never mimic
a canonical domain wall because, for a canonical wall, we always have that
\be
\phi'^2 = E(\phi) \ .
\ee
In the $P(X)$ DBI case, this would require the expressions on the 
right-hand side of (\ref{e:DBI1I}) and (\ref{e:DBIE}) to be equal.  A quick 
calculation shows that this can only happen if $V(\phi) = 0$, and hence the
$P(X)$ DBI wall can never mimic a canonical wall.

As another example, we consider a different $P(X)$ theory defined by
\be\label{e:X2}
P(X) = -\frac{1}{2} X + \alpha X^2 \ ,
\ee
where $\alpha$ is a real parameter with dimensions of [mass]$^{-4}$.  When
$X \ll \alpha$, this reduces to the canonical scalar field theory.
Following the
techniques used previously, we find that this theory
possesses a conserved quantity $J$ given by
\be
J = -\frac{1}{2} \phi'^2 + 3\alpha \phi'^4 + V(\phi) \ ,
\ee
where $V(\phi)$ is the potential associated with the theory.  One might
suppose that, since this theory is a deformation of the canonical one,
a deformation of the potential would suffice to mimic the canonical wall.
Again 
assuming that  the potential is 
positive semidefinite and  has discrete
zero-energy minima at $\phi = \phi_{\pm}$, with
$\phi_- < \phi_+$, so that $V(\phi_\pm) = 0$, and
assuming boundary conditions where
$\phi = \phi_\pm$ at $\pm\infty$, we find the first integral
\be
\phi'^2 = \frac{1 - \sqrt{1 - 48 \alpha V(\phi)}}{12\alpha} \ ,
\ee
whereas
\be
E(\phi) = \phi'^2 - 4 \alpha \phi'^4 \ .
\ee
Since $E(\phi) \ne \phi'^2$, we see that there is no choice of the potential
for which the theory defined by (\ref{e:X2}) mimics a canonical wall, so long
as $\alpha\ne 0$.

\subsubsection{Conditions for Doppelg\"anger Defects in More General Actions}

The discussion in the previous section does not imply the absence of
other  doppelg\"anger actions.  As we now
show, there are  infinitely many higher-derivative actions which 
can mimic canonical domain walls.  However, these other actions are ``rare" 
in the sense that they are technically non-generic in 
the space of all scalar field actions.  We  make this statement more
precise below.

Consider the family of scalar field actions which have
second-order equations of motion.  Such an action is defined by a
Lagrangian which is a function of 
 both $X = (\partial\phi)^2$ and $\phi$,
\be\label{e:genL}
L = L(X,\phi) \ ,
\ee
containing the much smaller family of $P(X)$
actions as a special case.
We denote the canonical action by $L_0$, so that
\be\label{e:canL}
L_0(X,\phi) = -\frac{1}{2} X - V(\phi) \ .
\ee
The conserved quantity for the general Lagrangian (\ref{e:genL}) is
given by
\be
J = 2 X \frac{\partial L}{\partial X} - L \ ,
\ee
whereas for the canonical action $J_0 = -X + V(\phi)$.  Without loss of generality we assume that the
domain wall boundary conditions are such that $J=0$ everywhere. This 
can always be enforced by shifting
$L$ by a constant $L(X,\phi) \to L(X,\phi) + c$, which does not affect 
the equations of motion and only shifts the zero point of the energy density.
For the canonical action, 
this implies that we can impose $V(\phi_{\rm min}) = 0$
for the global minima $\phi_{\rm min}$ of $V$.

What is required of a higher-derivative action so that it can mimic a
canonical scalar field action?  The first requirement is that both
actions must have the same field profile $\phi_0(z)$ as a solution to their
respective
equations of motion.  The second requirement is that the energy density of
this field profile be the same when evaluated using the Hamiltonians 
associated with 
their respective actions.  

We employ a geometrical construction to investigate these requirements.
Instead of viewing $L$ and $L_0$ as
functions, it is helpful to think of them as surfaces hovering over
the $(X,\phi)$ plane, with a height given by $L(X,\phi)$ or $L_0(X,\phi)$,
respectively.  These surfaces are referred to as the ``graphs" of the functions
 $L$ and $L_0$.  

We first consider the second requirement, that the field 
profile $\phi_0(z)$ has the same Hamiltonian energy densities in
the two theories.  Suppose we have already
established that the same field profile $\phi_0(z)$ is a solution to both
actions.  We denote by
$\phi_-$ the value of $\phi$ at $z = -\infty$ and by $\phi_+$ 
the value at $z = +\infty$ for this solution.
  The specified solution traces out a curve $C$
on the $(X,\phi)$ plane given in parametric form by
\be
C : z \mapsto ( X_0(z), \phi_0(z) ) \ .
\ee
Since the configurations are static, the energy density is simply
$-L$.  Hence we can satisfy the first requirement if and only if
\be\label{e:RQ2}
L(X,\phi) = L_0(X,\phi) \text{ on } C \ .
\ee
$L$ and $L_0$ need not agree everywhere, but they must agree when evaluated
on points on $C$.  Geometrically, (\ref{e:RQ2}) means that the 
graphs of $L$ and $L_0$ must intersect, and the projection of
this intersection on to the $(X,\phi)$ plane must contain $C$.

We next consider the first requirement,
 that the equations of motion for either action 
admit the specified field profile $\phi_0(z)$ as a solution.  We  assume
that $\phi_0(z)$ is a solution to the canonical theory, and derive the 
requirement that it also be a solution to $L$.  Recall that, for static
configurations, actions of 
the form (\ref{e:genL}) always admit a first integral obtained by solving the
equation $J =0$ for $\phi'^2$.  Hence, $J$ must vanish when evaluated on
the solution to the canonical theory.  That is, $\phi_0(z)$ will be a solution
to the higher-derivative scalar field theory if and only if
\be
2 X \frac{\partial L}{\partial X} - L = 2 X \frac{\partial L_0}{\partial X}  - L_0 \ \ \ \text{ on } C
\ee
which, using (\ref{e:RQ2}), yields
\be\label{e:RQ1}
\frac{\partial L}{\partial X} = \frac{\partial L_0}{\partial X}  \ \ \ \text{ on } C \ .
\ee
Hence, we require that the derivatives of $L$ and $L_0$ with respect to $X$
agree on $C$.  Note that we never need to match derivatives with respect to
$\phi$ - while $\partial L / \partial \phi$ does enter the equations of 
motion, it does not enter our conserved quantity and hence is not required
to find a solution.

We conclude that:

\begin{quotation}
\noindent \emph{An action $L(X,\phi)$  mimics a domain wall $\phi_0(z)$
of the canonical scalar field theory $L_0$ (that is, has the same
field profile and energy density) if and only if the graphs of
$L$ and $L_0$ intersect above the curve $C : z \mapsto (X_0(z),\phi_0(z))$
in the $(X,\phi)$ plane, and if 
$\partial L /\partial X = \partial L_0 /\partial X$ along the intersection.
}
\end{quotation}

\begin{figure}   
\includegraphics[width=4in]{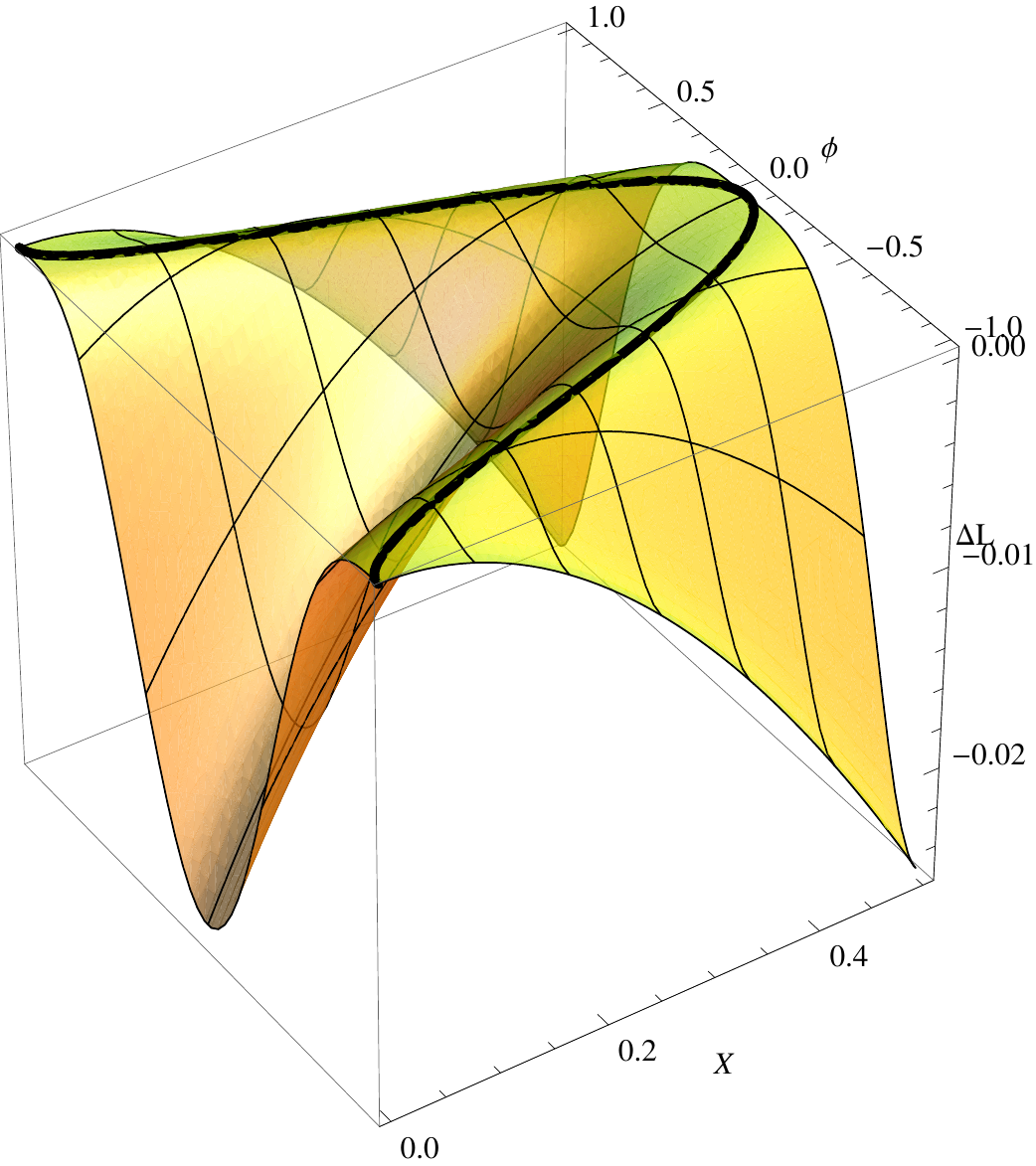}%
\caption{\label{f:unique} An illustration of the geometrical interpretation
of doppelg\"anger actions.  The graphs of $L(X,\phi)$ and 
$L_0(X,\phi)$ intersect along a single curve, whose 
projection on to the $(X,\phi)$ plane is the curve $C$ discussed in the 
text.
Here we plot the graph of $L_0 - L$ for the DBI action and the curve
$C$ (in black).
The intersection of the graphs of $L$ and $L_0$ is non-generic, since
the first derivatives of the $L-L_0$ surface vanish along $C$.}
\end{figure}  

This geometrical picture, when combined with the two constraints
(\ref{e:RQ2}) and (\ref{e:RQ1}), 
allows us to make a powerful statement about how
``rare" doppelg\"anger actions are.  The graphs of $L$ and $L_0$ are codimension-1
surfaces in the same three-dimensional space.  Hence, they will generically
intersect along a one-dimensional curve.  Thus, we should not be
surprised if two actions satisfy the constraint (\ref{e:RQ2}), which is
essentially the statement that the graphs intersect along a one-dimensional
curve.  However, two codimension-1 manifolds will generically intersect
``transversely" - the span of their tangent spaces will equal the tangent
space of the manifold at the intersection
 (${\bf R}^3$ in this case).  The condition
(\ref{e:RQ1}) implies that the graphs of $L$ and $L_0$ do not intersect
trasversely.  Thus, the existence of doppelg\"anger walls depends on constructing 
graphs in ${\bf R}^3$ which intersect non-generically. This geometrical interpretation is illustrated in Figure \ref{f:unique}, where
we have compared a canonical action with $V(\phi) = (1/4)(\phi^2-1)^2$ and
its doppelg\"anger Lagrangian.  

Another way of putting this result is that, given any function 
$\Delta L(X,\phi)$, such that
\be\label{e:DL}
\Delta L(X,\phi) = 0 \ \ \ \text{ on } C \quad \text{and} \quad
\frac{\partial\Delta L}{\partial X} = 0 \ \ \ \text{ on } C \ ,
\ee
then we can construct another action 
\be
L(X,\phi) = L_0(X,\phi) + \Delta L(X,\phi) \ ,
\ee
which will have the same domain wall solution as $L_0$.  Clearly there are
infinitely many functions $\Delta L$ satisfying (\ref{e:DL}), though they
are non-generic in the same sense as non-transversely intersecting
pairs of surfaces are non-generic.

\subsection{DNA Tests for Defects: Fluctuation spectra for Doppelg\"angers\label{ss:flux}}

The existence of doppelg\"anger defects raises the question of whether such objects are
merely a reparameterization of the original, canonical scalar field wall. As we shall demonstrate here,
the fluctuation spectra of the doppelg\"anger walls are
distinctly different from those of canonical walls.  Among other differences,
when the doppelg\"{a}nger walls are deeply in the DBI regime ($V_0/M^4$ large), 
they have far more bound
states than the canonical wall.  Since the fluctuation spectra are
different, the two theories cannot be reparameterizations of each other.

We find the action and equation of motion for the fluctuations by taking 
\be
\phi(t,z) = \phi_0(z) + \delta\phi(t,z) \ ,
\ee
where $\phi_0(z)$ is a static background 
solution to the equations of motion and $\delta\phi(t,z)$ the fluctuation.  
We then expand the Lagrangian 
to quadratic order in $\delta\phi$.  The  term linear in $\delta\phi$ 
vanishes since
$\phi_0(z)$ satisfies the equations of motion, and the purely quadratic
piece is of the form
\be
\delta_2L = A(z) \delta\dot\phi^2 + B(z) \delta\phi^2 + C(z) \delta\phi'^2 + D(z) \delta\phi \delta\phi' \ .
\ee
For the canonical action, $A = 1/2$, $B = -V''(\phi_0(z))/2$, $C = -1/2$, and 
$D = 0$.  For other cases, these coefficients depend on the particular
background solution $\phi_0(z)$ and on the specific action used.

Since the action is independent of $t$, different frequencies do not mix and
we can study an individual mode with frequency $\omega$ by taking
\be
\delta\phi(t,z) = e^{-i\omega t} \delta\phi(z) \ .
\ee
This leads to the quadratic action 
\be
\delta_2 L =  (\omega^2 A(z) + B(z) ) \delta\phi^2 + C(z) \delta\phi'^2 + D(z) \delta\phi \delta\phi' \ ,
\ee
yielding the equation of motion
\be\label{e:SLP}
\frac{C}{A} \delta\phi'' + \frac{C'}{A} \delta\phi' + 
\left[ \frac{D'-2B}{2A}\right] \delta\phi = \omega^2 \delta\phi \ .
\ee
Finding the energies of the flutuation modes amounts to finding values
of $\omega$ so that (\ref{e:SLP}) is satisfied by a normalizable
function $\delta\phi$.  

The problem (\ref{e:SLP}) is an eigenvalue 
problem of the Sturm-Liouville type.  Ideally, it would be in the form
of a Schr\"{o}dinger equation, which would allow us to readily identify 
free and bound states by analogy to the corresponding quantum mechanical
system.  Unfortunately, 
in general (\ref{e:SLP}) is not of Schr\"{o}dinger type,
thanks to the presence of the
$\delta\phi'$ term. However, in many interesting cases the quantity
\be\label{e:E0}
E_0 \equiv \frac{D'-2B}{2A}
\ee
tends to a constant far away from the wall.  Hence, evaluating (\ref{e:E0})
far away from the wall defines an analogue to the ``binding
energy" of various fluctuation modes.  We call modes with $\omega^2 < E_0$
the ``bound states," and modes with $\omega^2 > E_0$ ``free states."
This definition gives reasonable agreement with our expectations for bound
and free states, as we discuss below.

\begin{figure}   
\includegraphics[width=6.3in]{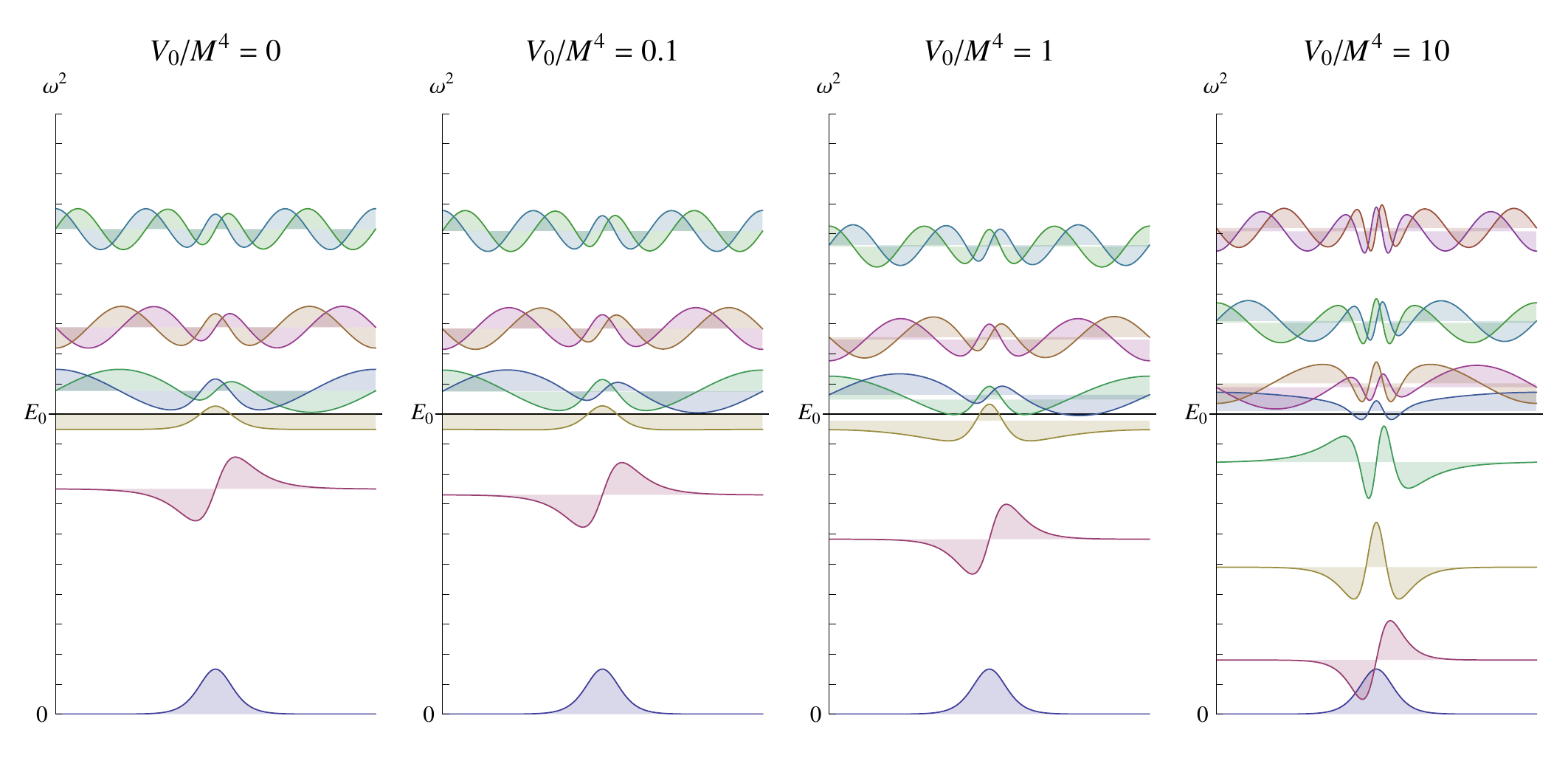}%
\caption{\label{f:flux}The lowest-lying fluctuation eigenmodes for 
various domain walls.  The vertical position of each eigenmode is the
eigenvalue $\omega^2$ normalized by the binding energy $E_0$.  Shown are
the spectra for a canonical scalar field wall with $V_0/M^4 = 0$ (leftmost panel) and then some of its doppelg\"{a}ngers with
 $V_0/M^4 = 0.1$, $1$,
and $10$ respectively.  As the ratio $V_0/M^4$ increases, the wall possesses
more bound states.  The lowest-lying state is identical for each wall,
reflecting the fact that these walls share a background field profile.}
\end{figure}

The eigenvalue problem (\ref{e:SLP}) can be solved numerically using a simple
finite element approach.  We have computed the lowest-lying 
eigenmodes for a canonical wall with potential
\be
V(\phi) = \frac{V_0}{4} \left( \phi^2 - \phi_0^2 \right)^2
\ee
and some of its doppelg\"anger walls, assuming periodic boundary conditions with 
periodicity much larger than the wall width.  Some  of these solutions are
shown in Figure \ref{f:flux}.  These figures show the energies $\omega^2$ 
of these
fluctuation modes, normalized to the binding energy $E_0$ defined in
(\ref{e:E0}), which is itself shown by the black horizontal line in the figure.
As can be seen, our definition of bound states is reasonable, since the
eigenmodes possess the properties one would expect of bound states (such
as compact support) when their energies are below $E_0$, and the 
properties of free states (such as oscillatory behavior) when their energies
are above $E_0$.  Since the eigenspectra are different, we can conclude
that the two theories, while possessing an identical background solution,
are in fact distinct theories.

The figures also show that there are many more bound states for the doppelg\"anger
wall when we increase the mass scale of the potential relative to the DBI 
scale.  These bound states are possible because the DBI action ``weights"
gradient energy much less in the interior of the domain wall, and hence 
even highly oscillatory fluctuation modes can remain as bound states.  Physically,
the presence of these bound states means that the doppelg\"anger wall possesses
additional oscillation modes which the canonical wall does not.

\section{$k$-strings}\label{s:kstrings}

It is natural to ask whether it is also possible to find doppelg\"angers of other
defect solutions, such as global strings or monopoles.  
This question is somewhat difficult to answer since higher codimension
defects are generally less analytically tractable than the domain wall.
In particular, the existence of the conserved quantity $J$ in the
codimension-one (domain wall) case allowed us to 
find the field profile and energy
density and construct a doppelg\"anger existence proof.
No analogous quantity is available for higher codimension defects,
such as global strings or monopoles.

In this section, we generalize the one-field DBI action to a two-field
system, and investigate some properties of the correponding global string
solutions.  Since we have no conserved quantity, we take a numerical
approach and directly integrate the equations of motion.  Using our
two-field DBI model, we find no doppelg\"anger global string solutions.
Nevertheless, since we cannot treat the two-field system analytically, we
cannot prove a `no-go' theorem and hence the existence of higher
codimension doppelg\"anger defects
remains an open question.

The canonical global string solution can be found by starting from the
action with two real scalar fields
\be\label{e:canStringS}
S = \int \left[-\frac{1}{2} (\partial \phi_1)^2 -\frac{1}{2} (\partial \phi_2)^2
- V(\phi_1,\phi_2)\right] \; \dd^4 x \ ,
\ee
where the potential $V(\phi_1,\phi_2)$ respects a global $O(2)$
symmetry, corresponding to rotations in the $(\phi_1,\phi_2)$ plane.
To study string solutions, we assume the field configuration is 
static and cylindrically
symmetric, employ polar coordinates $(r,\theta)$ in real space, and use the rotational symmetry to
decompose the fields in terms of new functions $\phi$ and $\Theta$ as
\be\label{e:red1}
\phi_1(r,\theta) = \phi(r) \cos \Theta(N\theta), \quad \phi_2(r,\theta)
 = \phi(r) \sin \Theta(N\theta) \ ,
\ee
where $N \in \mathbf{Z}$ is the winding number of the string.  Restricting ourselves
to strings of unit winding number $N=1$, the entire
action may then be written in terms of the single function $\phi(r)$.  The
equation of motion for this field is
\be
\phi'' + \frac{\phi'}{r} - \frac{\phi}{r^2} - \frac{\partial V}{\partial \phi} = 0 \ ,
\ee
where $\phi' = \partial\phi/\partial r$.
Given a potential $V(\phi)$ which admits a defect solution, that is, $V(0) \neq 0 $ and there exists $\phi_0 > 0$ such that $V(\phi_0)=0$ is a minimum, the string solution is subject to the boundary conditions that $\phi(0) = 0$ and $\phi\rightarrow \phi_0$ as $r\rightarrow \infty$.  It is then straightforward to solve for the string field profile using the relaxation method.

There are many multi-field
generalizations of the basic DBI kinetic term (\ref{e:DBI})
which appear in the literature.  Typically these generalizations reduce
to the usual DBI kinetic term when there is only a single field.  Based
on our experience with the doppelg\"anger solutions, the best-motivated generalization is analogous to (\ref{e:NG1}), 
based on a generalization of the Nambu-Goto action with 
two extra dimensions given by
\be\label{e:NG2}
S_{NG} = -  \int T(X) \sqrt{ - \det \left[ \eta_{MN}\frac{\partial X^M}{\partial x^\mu}
\frac{\partial X^N}{\partial x^\nu} \right] } \; \dd^4 x \ ,
\ee
where, as before, the tension $T$ is a function of the
embedding coordinates.  We depart from (\ref{e:NG1}) by taking
 six-dimensional embedding coordinates
$X^N$, with $N=0...5$ and
\be
X^N = x^N \; : \; N = 0,...3, \qquad X^4 = \phi_1(x^\mu) \ ,
\;\; X^5 = \phi_2(x^\mu) \ .
\ee
Hence, the four-dimensional theory contains two real
scalar fields $\phi_{1,2}$ with an $O(2)$ global symmetry.  With 
a suitable choice of tension $T(X)$, we can construct DBI generalizations
of the usual global string.

At this point, we can follow a similar procedure to that carried out in 
the case of the canonical global string.  The reduction of the fields
in the case of the unit winding number string proceeds exactly as before,
with the same decomposition defined by (\ref{e:red1}).
If we use this decomposition in (\ref{e:NG2}) we find
\be
S_{NG} = 2 \pi \int \left[ 
r - (1 +U(\phi))
\sqrt{ ( r^2 + \phi^2 ) ( 1  + \phi'^2 ) }
\right] \; \dd r \ ,
\ee
where, as before, we have rewritten $T = 1+U(\phi)$ and added a 
constant to the Lagrangian so that the energy is zero when $\phi'=0$ and
$U(\phi)=0$.
Note that there is no factor of $r$ next to the differential, since the
action (\ref{e:NG2}) already correctly accounts for the volume measure in
four dimensions.

To investigate whether doppelg\"anger strings can be constructed, we
assume a symmetry-breaking potential $U(\phi) = U_0 (\phi^2 - 1)^2$ 
in the DBI theory
and solve via the relaxation method for the DBI field profile. 
Given the field profile $\phi(r)$ of the DBI string, we solve 
numerically for the potential in the canonical scalar field theory
which gives the same field profile.  With this potential function, we
compute the energy density in the canonical theory.
In the examples we study,  we find that the energy densities are different in the two  theories.  Analogous
results hold if we match energy densities between the DBI and 
canonical theory -- we find the field profile does not match. 
Hence we do not find any doppelg\"anger defects.

\begin{figure}   
\includegraphics{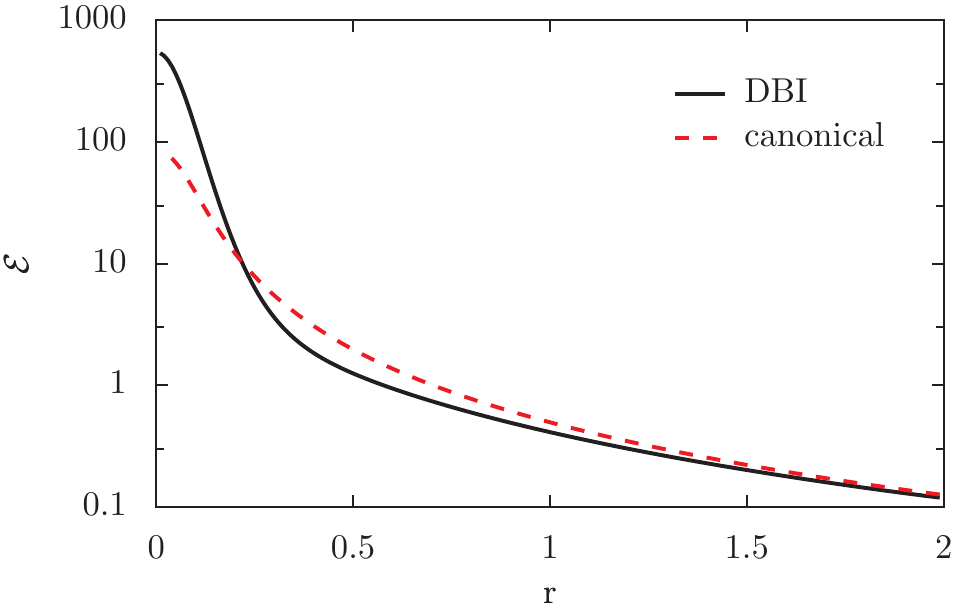}
\caption{\label{f:energy}Energy density as a function of radius for a DBI string and a canonical string with identical field profiles.  The DBI potential is given by $U(\phi) = 10 (\phi^2 - 1)^2$.}
\end{figure}

When taking the field  profiles to be equal, we can construct a potential such that the DBI field profile is a solution to the canonical equations of motion by integrating the canonical equation of motion for $\phi$, setting the potential to be $0$ at large $r$:
\be 
V(\phi) = \int_\phi^{\phi_0} \left( \tilde\phi'' + {\tilde\phi' \over r} - {\tilde\phi \over r ^2}  \right)
\;  \text{d} \tilde\phi
\ee
For the examples we have studied, this leads to a total energy density
which differs from the DBI energy density,  as shown in Figure \ref{f:energy}.

\begin{figure}   
\includegraphics{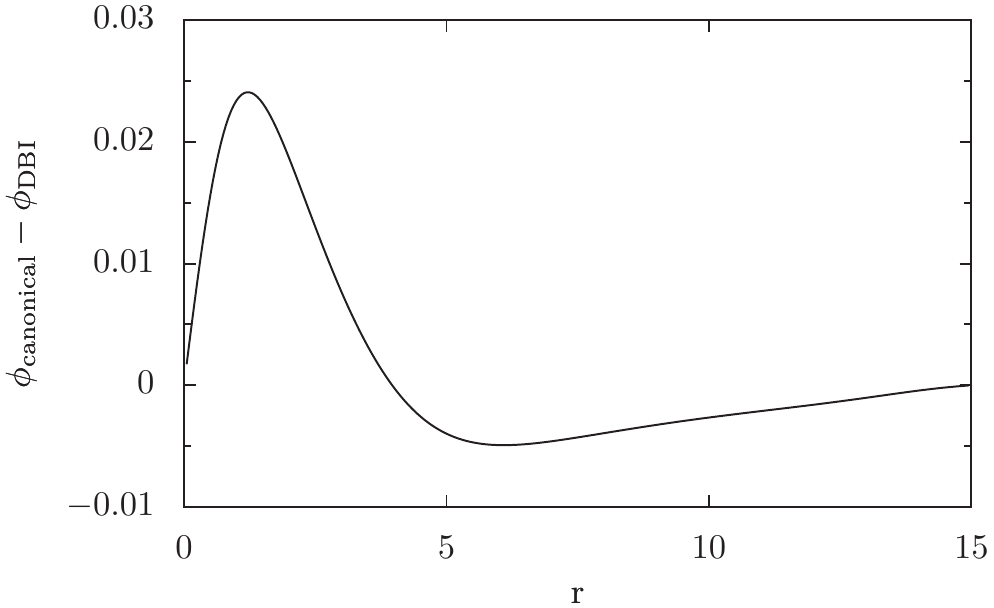}
\caption{\label{f:phi} The difference in field values for a DBI string and a canonical string with identical energy densities. The DBI potential is given by $U(\phi) = {1\over4} (\phi^2 - 1)^2$.}
\end{figure}

We also consider the case where the energy densities are constrained to be equal.  In this case, after solving for the field profile and energy density of the DBI string, we then similarly solve for the field profile of the canonical string while maintaining the canonical potential as $V = \mathcal{E}_{\rm DBI} - {1\over2} \left(\phi'^2_{\rm canonical} + {\phi^2_{\rm canonical} \over r^2} \right)$.  The results are shown in Figure \ref{f:phi}.

The two approaches, both constraining the field profiles to be equal and constraining the energy densities to be equal, yield a DBI string which is observably different from the canonical string for the examples we have
studied.  Thus we have found no examples of doppelg\"anger solutions
for cosmic strings.

\section{Discussion}\label{s:discussion}

Nonperturbative field configurations such as topological defects may be formed during phase transitions in the early universe, and their
interactions and dynamics can have significant effects on cosmic evolution. In the case of a scalar field with a canonical kinetic term, the behavior of such configurations has been understood for some time. The resulting constraints on the types and scales of symmetry breaking are well-understood, and the possibilities for interesting cosmological phenomena have been thoroughly investigated.

However, in recent years, particle physicists and cosmologists have become interested in non-canonical theories, such as those
that might drive $k$-inflation and $k$-essence. 
Ghost-free and stable examples of such theories can be 
constructed, and as such one
may take them seriously as microphysical models. Several authors have then studied the extent to which the properties of topological 
defects are modified by the presence of a more complicated kinetic term.

In this paper we have studied $k$-defect solutions to the DBI theory in some detail, discussing walls and strings, and clarifying the
existence criteria and the behavior of instantons in these theories. Furthermore, we have addressed the question of whether $k$-defects, and in particular $k$-walls and global $k$-strings, can mimic
canonical defects. We have demonstrated that given a classical theory with a canonical kinetic term and a spontaneously
broken symmetry with a vacuum manifold admitting domain wall solution, there exists a large family of general Lagrangians of 
the $P(\phi,X)$ form which admit domain wall solutions with the same field profiles and same energy per unit area. These doppelg\"anger defects
can mimic the field profile and energy density of canonical domain walls.
Nevertheless, we have also shown that the fluctuation spectrum of
a doppelg\"anger is different from its canonical counterpart, 
allowing one in principle to distinguish a canonical defect from its doppelg\"anger.

In the case of cosmic strings we have been unable to prove a similar result.
Despite investigating 
several examples for the potential function in the DBI theory,
we have been unable to find cases where there is a canonical theory which
results in a matching energy density and field profile.
However, since we have less analytic control in the case of defects of
higher codimension, we have not been able to prove a 'no-go' theorem.
Hence the existence of doppelg\"anger defects for strings or monopoles 
remains an open question.

\acknowledgments
We thank Adam Brown for discussions. This work is supported in part by National
Science Foundation grant PHY-0930521 and by Department
of Energy grant DE-FG05-95ER40893-A020.
MT is also supported by the Fay R. and Eugene L. Langberg Chair.

\end{document}